\documentclass[aps,preprint,preprintnumbers,nofootinbib,showpacs]{revtex4}
\usepackage{epsfig}

\def\beq{\begin{equation}}
\def\eeq{\end{equation}}
\def\beqa{\begin{eqnarray}}
\def\eeqa{\end{eqnarray}}

\def\ssc{\scriptscriptstyle}
\def\lsim{\mathrel{\raise.3ex\hbox{$<$\kern-.75em\lower1ex\hbox{$\sim$}}} }
\def\gsim{\mathrel{\raise.3ex\hbox{$>$\kern-.75em\lower1ex\hbox{$\sim$}}} }

\begin{document}
\preprint{{\vbox{\hbox{NCU-HEP-k047}
\hbox{Mar 2012}
}}}
\vspace*{.5in}

\title{Lie Group Contractions and Relativity Symmetries
\vspace*{.3in} }
\author{\bf Dai-Ning Cho and Otto C. W. Kong\vspace*{.2in}}
\email{otto@phy.ncu.edu.tw}
\affiliation{
Department of Physics and Center for Mathematics and Theoretical Physics,
National Central University,~Chung-li,~TAIWAN 32054.
 \vspace*{.8in}	}

\vspace*{.5in}
\begin{abstract}
With a more relaxed perspective on what constitutes a relativity symmetry
mathematically, we revisit the notion of possible relativity or kinematic
symmetries mutually connected through Lie algebra contractions.
We focus on the contractions of an $SO(m,n)$ symmetry as a relativity 
symmetry on an $m+n$ dimension geometric arena, which generalizes the
notion of spacetime, and discuss systematically contractions that reduce
the dimension one at a one, aiming at going one step beyond what has been
discussed in the literature. Our key results are five
different contractions of a Galilean-type symmetry $G(m,n)$ preserving a
symmetry of the same type at dimension $m+n-1$, {\it e.g.} a $G(m,n-1)$,
together with the coset space representations that correspond to 
the usual physical picture. Most of the  results are explicitly 
illustrated through the example of symmetries obtained 
from the contraction of $SO(2,4)$, which is the particular
case for our interest on the physics side as the proposed relativity
symmetry for ``quantum spacetime". The  contractions from
$G(1,3)$ may be relevant to real physics.
\end{abstract}

\maketitle

\section{introduction}
In the early 50's, In\"on\"u-Wigner contraction was introduced to understand
the structure between the symmetries for Einstein relativity and its low velocity
limit -- Galilean relativity \cite{IW}. The procedure and its generalizations
have since been established as a way of obtaining from semisimple Lie groups/algebras
interesting nilpotent ones. On the physics side, Lie algebra contraction has been
used to study plausible relativity symmetries, most noticeably in
Refs.\cite{con1,con2}.  Lie group deformation or stabilization \cite{CO}
is essentially the inverse process of such a contraction
\footnote{
For more discussions of algebra deformations and contractions in more
generic settings, see Refs.\cite{FM,BHRS}.}.
In particular, deformation of Einstein special relativity, or the 
Poincar\'e $ISO(1,3)$ symmetry
\footnote{
Here in our discussions, we are focusing only on the continuous, or so-called
homogenous, part of the symmetries, leaving aside plausible discrete symmetries
like parity and time reversal.},
was introduced as a mean to go beyond Einstein relativity to a new relativity
incorporating the quantum scale, as inspired by the early work of Snyder \cite{S}.
This perspective of a Quantum Relativity picture for the `quantum spacetime' is
the main idea behind an area of recent research \cite{A}. The interest in such
deformed special relativities was mostly rekindled by a series of papers around the
beginning of the century \cite{dsr}. We depart from most of the other works in
the literature by focusing on a simple picture within Lie group/algebra, as
advocated in Ref.\cite{CO}. Two important new ideas are introduce when putting
the mathematics into the suggested physics framework. The first
one is the linear (coset space) realization picture \cite{023}, like the Minkowski
spacetime realization of $ISO(1,3)$ symmetry. The latter, when applied to the
new relativity symmetries, dictates the radical idea of the `quantum spacetime'
being more than spacetime. One obtains a pseudo-Euclidean space with
new coordinates, the physical meaning of which is neither space nor time. Our
beginning exploration of the role of such coordinates in physics already shows
some interesting results \cite{023,030,036,037}. The second important idea we
introduced is that the quantum scale has to be incorporated into the symmetry
structure through two parameters instead of just one \cite{030,031}. The
two parameters, together with speed of light $c$, essentially give all the
fundamental constants one would think relevant to quantum spacetime, namely
$c$, the Newton constant $G$, and the quantum $\hbar$.

In Ref.\cite{036}, we introduced new relativities called Poincar\'e-Snyder and
Snyder relativities. These are supposed to be intermediate relativities between
Einstein relativity and the Quantum Relativity we identified in Ref.\cite{030}.
We have explored the classical and quantum mechanics of the Poincar\'e-Snyder
relativity of a $G(1,3)$ symmetry, with some success \cite{036,037}. The
current study is an out-growth from that work, aiming both to clarify the
relation between the mathematics and physics of all the possibly relevant
relativity symmetries under the framework. We also explore the somewhat boarder 
mathematics picture in the interest of generic mathematical physics. The strategy 
here is to analyze symmetry contractions starting from $SO(m,n)$ with 
$SO(2,4)$ as the specific one of interest for our physics program.

It is interesting to note that on the mathematical physics side, there has been
interesting recent developments on the topic of symmetry (algebra) contractions
\cite{glc,gdc,ck}. In particular, Ref.\cite{ck} provides a nice framework for
describing a large set of interesting contractions within the framework of
the $SO(m,n)$ symmetries.

In the next section, we discuss our perspective on what may constitute a
relativity symmetry, which sets the stage for the Lie algebras we choose to focus on.
The perspective represents a plausible modification or enrichment of the concept of
relativity symmetry, as compared to the traditional one, as given in the classic
1968 analysis \cite{con1}.
In Sec.III, we discuss the mathematics of symmetries through applying the contractions.
We basically use only the simple contractions of In\"on\"u-Wigner. We will however
give some comments on the relation between our results and the other contraction
pictures in the appendix. A first look at the physics picture of some plausible
relativity symmetries connected to our theme in the Quantum Relativity effort
will be discussed in Sec.IV, after which we conclude the paper.

\section{relativity symmetry}
We are interested in group/algebra symmetry which could be the relativity symmetry in physics.
That is to say, we are interested in algebras or groups which maybe considered as the
isometry of some `classical' geometric arena, such as the familiar three dimensional
space or four dimensional spacetime, for the description of fundamental physics.
\footnote{
The perspective may seem to be too conservative, excluding important new
mathematics plausibly relevant for quantum spacetime, for examples quantum
groups and noncommutative geometry. However, we think
about it as a focusing on the more familiar tools. Our analysis of the Quantum
Relativity apparently indicates that the classical (non-spacetime) geometric
picture with Lie symmetry may offer an alternative way to look a (quantum)
noncommutative spacetime geometry \cite{023,030,031}. The latter maybe somewhat
analog to the duality of the intrinsic and extrinsic description of
non-Euclidean geometry -- a curved space maybe described as a submanifold
of a flat Euclidean one. Such a classical non-spacetime description of the
`quantum spacetime', we believe, works at least at the `special' Quantum
Relativity limit -- one with essentially zero gravity.
}
We take a rotation-type $SO(m,n)$ symmetry as a standard, indispensable, part. The
rationale behind it is dictated by the physical picture of the isotropic arena having
real coordinates. Our Quantum Relativity
symmetry is $SO(2,4)$ \cite{030}. $SO(m,n)$ as a semisimple symmetry is
stable against deformations. Hence, studying symmetry contractions starting
from $SO(m,n)$ is a sensible strategy. The symmetry acts naturally on a
$(m+n)$ dimensional classical geometry, or a hypersurface of constant `radius'
inside it. Such a geometric realization is what we will stick to in all our
considerations of any one of the related mathematical symmetries regarded 
as a relativity symmetry. We mostly use the phrase `dimension of the 
relativity' to mean the dimension of such a geometry here. We emphasize 
again that the physical interpretation of classical geometric fundamental 
arena should not be restricted to the usual spacetime one.

Usually, we like to consider the arena as admitting also (coordinate) translation 
symmetries, though our Quantum Relativity is rather a counter example in this 
respect. A pure $SO(m,n)$ symmetry as a relativity/kinematic symmetry is
actually familiar \cite{con1}, under the picture of curved `spacetime'. There
has also been interesting recent developments in some particle dynamics picture
of de-Sitter special relativity \cite{dssr}. With the commuting translations
added, that gives rises to an $ISO(m,n)$ symmetry. A further kind of nontrivial
symmetry in a relativity is given by the example of the Galilean boosts. We
call boosts here all such symmetries characterized by the structure as
translations depending on a parameter external to the arena, for instance the
Galilean time. The Galilean (velocity) boosts together with the time translation
supplemented to an $ISO(0,3)$ gives the Galilei group $G(0,3)$. From the
mathematics point of view, the boosts can be defined through the specific
commutation relationship between their generators and those of the rotations
and translations. In that sense, boosts are boosts only relative to a set
of translations and a `Hamiltonian' generator exemplified by the generator of
time translation in $G(0,3)$. Hence the so-called Lorentz boosts are no boosts;
they are spacetime rotations. Under this framework, we have introduced
the $G(1,3)$ symmetry for what we called Poincar\'e-Snyder relativity
\cite{036,037}. The latter as an extension of the familiar $ISO(1,3)$ is
a descendant from the $SO(2,4)$ quantum relativity symmetry \cite{030} via
contractions through an $ISO(1,4)$. Those symmetries and others connected
to them through symmetry algebra contractions are the ones we will
discuss in this paper.

The key results of the paper are symmetries obtained from contractions
of $G(m,n)$ preserving a subgroup of the same type of dimension
$m+n-1$, illustrated by the case of $G(1,3)$, together with the contracted
coset space representations corresponding to the usual relativity picture
of the Newtonian/Galilean space and Einstein/Minkowski spacetime. The
physical picture maybe relevant to nature under the background
perspective of our Quantum Relativity framework.

\section{\boldmath Relativity Symmetry Contractions from  $ISO(m,n)$ and $G(m,n)$.}
A generic picture on contractions from $SO(m,n)$ to $ISO(m-1,n)$ or
$ISO(m,n-1)$ as well as contractions from $ISO(m,n)$ to $G(m-1,n)$ or
$G(m,n-1)$ can be find in Gilmore's book\cite{G}. For an illustration
within our perspective, let us sketch the contraction of an $ISO(m,n)$
type symmetry to one of $G(m,n)$ type, using $ISO(1,4)$ to $G(1,3)$
as an explicit example. It starts by
picking a subalgebra to be preserved. We have in mind going
from a possible relativity symmetry on a $(m+n)$ dimensional geometry
to one of a $(m+n-1)$ dimensional geometry. The one dimension `removed'
may be taken to be one of time-like ($-$) or space-like (+) geometric
signature. We take the latter choice for $ISO(1,4)$ to $G(1,3)$.
That means we pick $SO(m,n-1)$, {\it i.e.} $SO(1,3)$ in this case, as part
of the subalgebra. Unlike the simplest case of contraction from $SO(m,n)$,
$SO(m,n-1)$ is not a maximal subalgebra of $ISO(m,n)$. The nonzero
commutators among generators of $ISO(1,4)$ can be given as
\beqa
[J_{\mu\nu},J_{\lambda\rho}]  &=&
-(\eta_{\nu\lambda}J_{\mu\rho}-\eta_{\mu\lambda}J_{\nu\rho}
+\eta_{\mu\rho}J_{\nu\lambda}-\eta_{\nu\rho}J_{\mu\lambda}) \;,
\nonumber \\ &&
[J_{\mu\nu},J_{{\ssc 4}\rho}] = -(\eta_{\nu\rho}J_{{\ssc 4}\mu}-\eta_{\mu\rho}J_{{\ssc 4}\nu})\;,
 \nonumber \\ &&
[J_{\mu\nu},P_{\rho}] = -(\eta_{\nu\rho}P_{\mu}-\eta_{\mu\rho}P_{\nu})\;,
\nonumber \\   &&
[P_{\ssc 4},J_{{\ssc 4}\mu}] = -P_{\mu}\;,
\nonumber \\  &&
[P_{\mu}, J_{{\ssc 4}\nu}] = \eta_{\mu\nu} P_{\ssc 4}\;,
\nonumber \\ &&
[J_{{\ssc 4}\mu},J_{{\ssc 4}\nu}] = J_{\mu\nu}\;,
\label{i14}
\eeqa
where the index $4$ denotes the dimension with  space-like geometric
signature that is singled out. We then take $J_{{\ssc 4}\mu}$ to be among
the generators to go through the algebra transformation the singular
limit of which gives the contraction. The set of $J_{\mu\nu}$ gives
the preserved $SO(1,3)$ symmetry. To keep the resulting symmetry as a $1+3$
dimensional relativity symmetry, the $P_\mu$ generators obviously have
to be treated in the same way. If we take $P_{\ssc 4}$ with $J_{\mu\nu}$
and redefine the generators by the one parameter transformation
\begin{table}[t]
 \caption{\footnotesize            A schematic presentation of possible $1+3$
dimensional relativity symmetries from contractions of $ISO(1,4)$. For each case,
we listed the generators to be transformed, {\it i.e.} taken through the singular
limit of a transformation as illustrated in Eq.(\ref{kap}). In the contracted
symmetries, the $J_{{\ssc 4}\mu}$ or rather the $K^\prime_\mu$ generators can be taken
as a kind of boosts, or translations dependent on an extra parameter and the $P_{\ssc 4}$
generator is translation in the parameter, in the picture of $G(1,3)$. The extra parameter
can be taken as an `evolution' parameter with $P_{\ssc 4}$ corresponding to the `Hamiltonian'
of the `evolution' in a canonical formulation \cite{037}. Mathematically as defined in
the text, $K^\prime_\mu$  are boost generators in relation to the translations $P_\mu$
and the other generators in $G(1,3)$. There are no boosts in $S(1,3)$ and $S(1,3)$.
At the upper level of say $SO(2,4)$, all five $P_{\ssc A}$ will be $J_{5{\ssc\! A}}$
with the standard nonzero commutators among them. Note that all three cases have the
same $ISO(1,3)$ subalgebra, $J_{\mu\nu}$ and $P_{\mu}$ part, inside the resulting algebra.
}
\label{table1}
\begin{center}
\begin{tabular}{||c||c||c|c|c||c||}    \hline\hline
		&	&\multicolumn{3}{|c||}{Contracted	Symmetries } &\\ \hline
		& \ $ISO(1,4)$\ \ &\ $G(1,3)$\ \ 	&\ $C(1,3)$\ \ 	&\ $S(1,3)$	&\ \ \\ \hline
$[J_{\mu\nu},J_{\lambda\rho}]=$		&$J_{\mu\lambda}$	&$J_{\mu\lambda}$	& $J_{\mu\lambda}$	& $J_{\mu\lambda}$		&\\
$[J_{\mu\nu},J_{{\ssc 4}\mu}]=$		&$J_{{\ssc 4}\nu}$	&$J_{{\ssc 4}\nu}=\bar{\kappa} K^\prime_{\nu}$	&$J_{{\ssc 4}\nu}=\bar{\kappa} K^\prime_{\nu}$	 &$J_{{\ssc 4}\nu}=\bar{\kappa} K^\prime_{\nu}$	&  $[J_{\mu\nu},K^\prime_{\mu}]=K^\prime_{\nu}$\\
$[J_{\mu\nu},P_{\mu}]=$		&$P_{\nu}$	&$P_{\nu}=\bar{\kappa} P^\prime_{\nu}$ 	&$P_{\nu}$ 	&$P_{\nu}=\bar{\kappa} P^\prime_{\nu}$ 	& $[J_{\mu\nu},P^\prime_{\mu}]=P^\prime_{\nu}$\\
$[J_{{\ssc 4}\mu},P_{\ssc 4}]=$		&$P_{\mu}$	&$P_{\mu}=\bar{\kappa} P^\prime_{\mu}$ 	&0 	&0	&	$[K^\prime_{\mu},P_{\ssc 4}]= P^\prime_{\mu}$\\
$[J_{{\ssc 4}\mu},P_{\mu}]=$		&$P_{\ssc 4}$	&0	&$P_{\ssc 4}=\bar{\kappa} P^\prime_{\ssc 4}$	&0	& $[K^\prime_{\mu},P_{\mu}]= P^\prime_{\ssc 4}$	\\
$[J_{{\ssc 4}\mu},J_{{\ssc 4}\nu}]=$		&$J_{\mu\nu}$	&0	&0	&0	
        & 	\\
\hline
	&	generators transformed &$J_{{\ssc 4}\mu}, P_{\mu}$	&$J_{{\ssc 4}\mu}, P_{\ssc 4}$	&$J_{{\ssc 4}\mu}, P_{\mu}, P_{\ssc 4}$ &\\
\hline\hline
\end{tabular}
\vspace*{.2in}\hrule
\end{center}
\end{table}
\beq \label{kap}
\left(\begin{array}{c}
J_{\mu\nu}\\
P_{\ssc 4}\\
K^\prime_{\mu} \\  
P^\prime_{\mu}  \\  
\end{array}\right):=
\left(\begin{array}{cccc}
I &0 &0 &0\\
0 &I &0 &0\\
0 &0 &\frac{1}{{\bar{\kappa}}}I &0\\
0 &0 &0 &\frac{1}{{\bar{\kappa}}}I
\end{array}\right)
\left(\begin{array}{c}
J_{\mu\nu}\\
P_{\ssc 4}\\
J_{{\ssc 4}\mu}\\
P_{\mu}
\end{array}\right) \;,
\eeq
the singular limit of ${\bar{\kappa}} \rightarrow \infty$ gives the $G(1,3)$
algebra. The latter as a relativity symmetry was introduced together with a
physical picture in Refs.\cite{036,037}. The parameter ${\bar{\kappa}}$ will
be a physical constant with a similar role to $c$, the speed of light.

In the place of $P_{\ssc 4}$, one may choose the $P_\mu$ or simply nothing 
to be included into the subalgebra with $J_{\mu\nu}$. In each case taking the 
rest with $J_{{\ssc 4}\mu}$ through the singular transformation gives a
contraction. It is easy to see that the resulting symmetries are $C(1,3)$ and
$S(1,3)$, respectively, as given in Table~\ref{table1}. The table only presents
a schematic form of the commutators among the generators but should be enough
to illustrate the algebras. A contraction simply keeps some commutators
unchanged while trivializing some others as shown. The scheme obviously
works for any $(m,n)$.
Similarly, contractions from $ISO(m,n)$ keeping $SO(m-1,n)$, that is
`removing' a dimension of time-like geometric signature, gives rise to
$G(m-1,n)$, $C(m-1,n)$, and $S(m-1,n)$.
The $G(0,3)$, $C(0,3)$, and $S(0,3)$ relativities are the Galilean, Carroll,
and static relativities in the literature (for example, Refs.\cite{con1,con2}).
Our notation of $C(m,n)$, and $S(m,n)$ follows from there.
\begin{figure}[t]
\caption{An illustration of some of the contraction paths relating some
relativity symmetries of difference up to two in dimensions.
}
\hspace{ 0.0cm}
\vspace{-0.5cm}
\centerline{\epsfig{figure=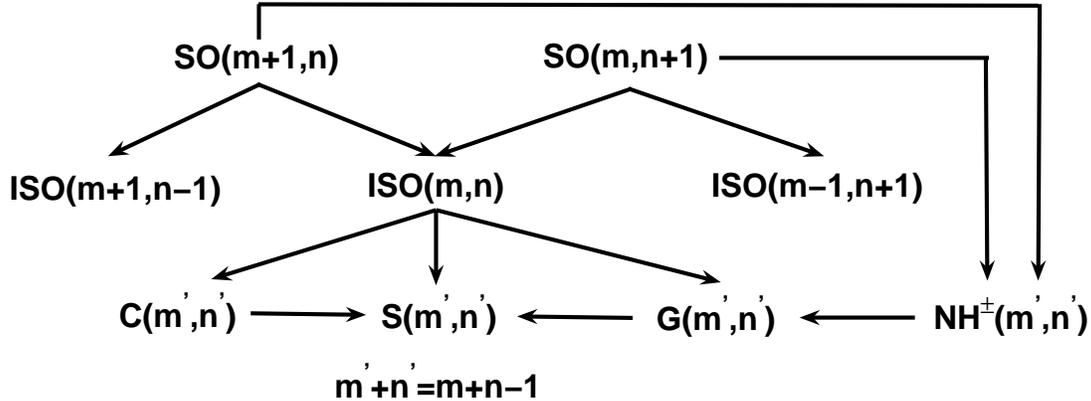,height=6.0cm,width=15.0cm}}
\vspace*{.2in}
\hrule{}
\label{fig1}
\end{figure}

The scheme above can be taken to define $G(m,n)$, $C(m,n)$, and $S(m,n)$
symmetries on any $(m+n)$ dimensional geometry.  It is
also easy to see simple contractions from $G(m,n)$ or $C(m,n)$ to $S(m,n)$
at the same dimension. For our illustrative case in the $ISO(1,4)$ notation,
the latter are achieved by taking $J_{{\ssc 4}\mu}$ and $P_{\ssc 4}$ through
another singular transformation in the first case, and $J_{{\ssc 4}\mu}$ and
$P_{\mu}$ in the second. We illustrate the above contraction paths
connecting the symmetries in Fig.1. 

We note in passing that if one consider contractions reducing the dimension
by two directly, it is possible to get a symmetry beyond what one can get by
taking a two step contraction each reducing the dimension by one.
An interesting example is given by the $N^+(0,3)$ Newton-Hooke case from
$SO(1,4)$ \cite{con1,con2} by-passing $ISO(1,3)$. To match the structure
of the analogous $N^+(1,3)$ versus those in Table~\ref{table1}, it has the
same commutator set as $G(1,3)$ with an extra nonzero $[P_\mu, P_{\ssc 4}]$.
Using $P_{\ssc 4}$, which is $J_{\ssc 54}$ at the $SO(2,4)$ level, gives
$[P_\mu, P_{\ssc 4}]=J_{{\ssc 4}\mu}$, {\it i.e.} the commutator of the 
translations with the Hamiltonian gives the corresponding boost generators. 
So, the mathematical role of $P_\mu$ and $J_{{\ssc 4}\mu}$ are actually
symmetrical in $N^+(1,3)$. They are both symmetries of the translation type
mathematically. The subalgebra generated by $J_{\ssc 54}$ ($P_{\ssc 4}$) and
$J_{\mu\nu}$ is preserved when contracting from $SO(2,4)$ directly,
{\it i.e.} the generators taken through the singular transformations are
$J_{5\mu}$ ($P_\mu$) and $J_{{\ssc 4}\mu}$.  Note that taking out only
$P_\mu$($J_{5\mu}$) without $P_{\ssc 4}$ ($J_{5\!4}$) and $J_{{\ssc 4}\mu}$ does
not leave a closed subalgebra. Similarly, one can project getting $N^+(m',n')$
symmetries from contractions of an $SO(m,n)$ with $m+n=m'+n'+2$.
And it can be further contracted to $G(m',n')$ or $S(m',n')$ at the same
dimension (see also Fig.1), by again taking $J_{{\ssc 4}\mu}$ and $P_{\ssc 4}$, or
$J_{{\ssc 4}\mu}$, $P_{\ssc 4}$, and $P_\mu$ through another singular transformation,
respectively, in our illustrative case for instance. It can be seen easily
that the $SO(m,n)$ can also be directly contracted to an $S(m',n')$
with $m+n=m'+n'+2$, but not $G(m',n')$ or $C(m',n')$.

Next, we look at contractions from the $G(m,n)$, reducing the dimension by
one. This could be useful, for example, in thinking about the low-velocity
limit of the Poincar\'e-Snyder $G(1,3)$ relativity we studied in
Refs.\cite{036,037}. We will explore that physical picture in the next section.
Here, we use $G(1,3)$ for an explicit illustrations of some the mathematics for
contraction from a $G(m,n)$, focusing of reducing to a symmetry on a geometry of
one less dimension with $G(m-1,n)$ or $G(m,n-1)$ to be preserved. This is
to maintain similar features in the contraction sequence from the types of
algebras $SO(m,n)$ to $ISO(m,n)$ to $G(m,n)$. In particular, we preserve the
$G(0,3)$ subalgebra in our example.

The (nonzero) commutators of $G(m,n)$ are given explicitly here in the notation
with boosts $K^\prime_{\mu}$ and `Hamiltonian' $H^\prime \equiv P_{\ssc 4}$,
splitting the time-like dimension from the three space-like ones :
\beqa
&& [J_{ij},J_{kl}]=-(\eta_{jk}J_{il}-\eta_{ik}J_{jl}+\eta_{il}J_{jk}-\eta_{jl}J_{ik}) \;,
\nonumber \\  &&
[J_{ij},J_{{\ssc 0}k}] = -(\eta_{jk}J_{{\ssc 0}i}-\eta_{ik}J_{{\ssc 0}j}) \;,   
\qquad
[J_{{\ssc 0}i},J_{{\ssc 0}j}] = -J_{ij} \;,   \nonumber \\  &&
[J_{ij},P^\prime_{k}] = -(\eta_{jk}P^\prime_{i}-\eta_{ik}P^\prime_{j}) \;,   \nonumber \\  &&
[J_{{\ssc 0}i},P^\prime_{\ssc 0}] = -P^\prime_{i} \;,   
\qquad
[J_{{\ssc 0}i},P^\prime_{i}] = -P^\prime_{\ssc 0} \;,   \nonumber \\  &&
[J_{ij},K'_{k}] = -(\eta_{jk}K'_{i}-\eta_{ik}K'_{j}) \;,   \nonumber \\  &&
[J_{{\ssc 0}i},K'_{\ssc 0}] =  -K'_{i} \;,   
\qquad
[J_{{\ssc 0}i},K'_{i}] = -K'_{\ssc 0} \;,   \nonumber \\  &&
[H',K'_{\ssc 0}] = -P^\prime_{\ssc 0} \;,   
\qquad
[H',K'_{i}] = -P^\prime_{i}\; .
\eeqa
Denote the subalgebra to be preserved in the contraction process by {\boldmath ${h}$}
and the complementary vector-subspace  {\boldmath ${p}$}. Recall that generators of
{\boldmath ${p}$} are to be taken to the limit of the singular transformation.
Our first task is to look at choices of the {\boldmath ${h}$} and {\boldmath ${p}$}
splitting. The $J_{ij}$ generators of the $SO(0,3)$ subalgebra is certainly
what we want to keep in going from $1+3$ dimension to $0+3$ dimension.
Notice that inside the $G(1,3)$, there is an $ISO(1,3)$ subalgebra. From the
above analysis, we can see that among the generators of the latter, keeping
$P^\prime_{\ssc 0}$ or $P^\prime_i$ or nothing together with the $J_{ij}$ will
contract the $ISO(1,3)$ to $G(0,3)$, $C(0,3)$, or $S(0,3)$, respectively. As said,
we stick here to the case getting $G(0,3)$.
\footnote{
It is interesting to note that while the $SO(m,n)$ contractions preserve
either an $SO(m-1,n)$ or an $SO(m,n-1)$, the $ISO(m,n)$ contractions still
preserve either an $ISO(m-1,n)$ or an $ISO(m,n-1)$ (see Table~\ref{table1} for
example). Contraction from a $G(m,n)$, however, may not preserve a
$G(m-1,n)$ or a $G(m,n-1)$ in general. }
Keeping $P^\prime_{\ssc 0}$ with $J_{ij}$ as the minimal set of generators for
{\boldmath ${h}$} and $J_{{\ssc 0}i}$ and $P^\prime_i$ in {\boldmath ${p}$}, one
can see that there are six possible choices of the {\boldmath ${h}$} and
{\boldmath ${p}$} splitting, as follows:
\[
\begin{array}{c|lrcc}
\mbox{Case} & \multicolumn{1}{c}{\mbox{\boldmath ${h}$}} & \multicolumn{3}{c}{\mbox{\boldmath ${p}$}} \\ \hline
A &{\ J_{ij},P^\prime_{\ssc 0}} &{H',K'_{\ssc 0},K'_{i},P^\prime_{i},J_{{\ssc 0}i}}
 & \rightarrow & {H'',K''_{\ssc 0},K''_{i},P_{i},K_i}\\
B &{\ J_{ij},P^\prime_{\ssc 0},H'} &{K'_{\ssc 0},K'_{i},P^\prime_{i},J_{{\ssc 0}i}}
 & \rightarrow & {K''_{\ssc 0},K''_{i},P_{i},K_i}\\
C &{\ J_{ij},P^\prime_{\ssc 0},K'_{\ssc 0}} &{H',K'_{i},P^\prime_{i},J_{{\ssc 0}i}}
 & \rightarrow & {H'',K''_{i},P_{i},K_i}\\
D &{\ J_{ij},P^\prime_{\ssc 0},H',K'_{\ssc 0}} &{K'_{i},P^\prime_{i},J_{{\ssc 0}i}}
 & \rightarrow & {K''_{i},P_{i},K_i}\\
E &{\ J_{ij},P^\prime_{\ssc 0},K'_{i}} &{H',K'_{\ssc 0},P^\prime_{i},J_{{\ssc 0}i}} & \rightarrow & {H'',K''_{\ssc 0},P_{i},K_i}\\
F &{\ J_{ij},P^\prime_{\ssc 0},K'_{\ssc 0},K'_{i}} &{H',P^\prime_{i},J_{{\ssc 0}i}}
 & \rightarrow & {H'',P_{i},K_i}
\end{array}
\]
where the new set of generators in {\boldmath ${p}$} are given by the original
set times ${1}/{c}$ with the singular limit $c \to \infty$. We skip the 
tedious mathematical details and give a schematic presentation of the results 
in Table~\ref{table2}. The cases A to F, as defined above are listed, with the 
resulting symmetries given the names or notations shown which are to be explained 
in the discussion below. The $(3)$ as in  $G_X(3)$ for all cases is a suppressed
form for $(0,3)$, indicating the $SO(0,3)$ subalgebra. A similar analysis applying
to $G(m,n)$ with the condition that a lower dimensional $C(m,n)$ or $S(m,n)$,
instead of $G(m,n)$, being the subalgebra to be preserved, can certainly be
performed to yield further alternatives.

\begin{table}[t]
 \caption{\footnotesize            A schematic presentation of
possible $0+3$ dimensional relativity symmetries from the contraction of $G(1,3)$.
Having the Galilean symmetry $G(3)$ [$\equiv G(0,3)$] as a subgroup is taken as
a criterion. The six different results of symmetry contractions follow from the
discussion in the text. As in the case of Table~\ref{table1}, new generators like
$K_i$, $P_i$, $K''_i$, ... etc.  shown in the last columns are to be used in the
contracted symmetries. $P^\prime_{\ssc 0}$ is to be renamed as $H$.
}
\label{table2}
\begin{center}
\begin{tabular}{||c||c|c|c|c|c|c|c||c|c||}    \hline\hline
	& Poincar\'e-Snyder	&F 	&E 	&D 	& C		& B	& A	& Galilean & \\
	& $G(1,3)$ &$G_{B}(3)$	&$G_{C}(3)$	&$G_{D}(3)$	& $G_{T}(3)$	& $G_{B}(3)$  	&  $G_{S}(3)$	&  $G(3)$ & \\
\hline\hline
$[J_{ij},J_{kl}]=$		&$J_{ik}$	&$J_{ik}$	&$J_{ik}$	&$J_{ik}$	&$J_{ik}$	&$J_{ik}$	&$J_{ik}$	&$J_{ik}$ & \\
$[J_{ij},J_{{\ssc 0}i}]=$		&$J_{{\ssc 0}j}$	&$J_{{\ssc 0}j}$	&$J_{{\ssc 0}j}$ 	&$J_{{\ssc 0}j}$ 	&$J_{{\ssc 0}j}$ 	&$J_{{\ssc 0}j}$ 	 &$J_{{\ssc 0}j}$ 	&$[J_{ij},K_{i}]=K_j$ & $K_{j}=\frac{1}{c}J_{{\ssc 0}j}$\\
$[J_{ij},P^\prime_{i}]=$		&$P^\prime_{j}$	&$P^\prime_{j}$	&$P^\prime_{j}$ 	&$P^\prime_{j}$ 	&$P^\prime_{j}$ 	&$P^\prime_{j}$ 	 &$P^\prime_{j}$ 	&$[J_{ij},P_{i}]=P_j$ & $P_{j}=\frac{1}{c}P^\prime_{j}$\\
$[J_{{\ssc 0}i},P^\prime_{\ssc 0}]=$		&$-P^\prime_{i}$	&$-P^\prime_{i}$	&$-P^\prime_{i}$ 	&$-P^\prime_{i}$ 	&$-P^\prime_{i}$ 	 &-$P^\prime_{i}$ 	&$-P^\prime_{i}$ 	&$[K_i,H]=P_i$ & $H=P^\prime_{\ssc 0}$\\
$[J_{{\ssc 0}i},J_{{\ssc 0}j}]=$		&$-J_{ij}$	&0	&0	&0	&0	&0	&0	& $[K_i,K_j]=0$ & \\
$[J_{{\ssc 0}i},P^\prime_{i}]=$		&$-P^\prime_{\ssc 0}$	&0	&0	&0	&0 	&0 	&0 	& $[K_i,P_j]=0$ & \\
\hline
$[J_{ij},K'_{i}]=$		&$K'_{j}$	&$K'_{j}$	&$K'_{j}$	& $K'_{j}$	&$K'_{j}$ 	&$K'_{j}$ 	&$K'_{j}$ 	
& \multicolumn{2}{c||}{ $[J_{ij},K''_{i}]=K''_j$, $K''_i=\frac{1}{c}K'_{i}$}  \\
& & \multicolumn{6}{c||}{ \big($K''_{\ssc 0}= \frac{1}{c}K'_{\ssc 0}$ for A, B, E \big)}  & \multicolumn{2}{l||}{\ for A, B, C, D}\\
$[J_{{\ssc 0}i},K'_{i}]=$		&$-K'_{\ssc 0}$	&0	&$-K'_{\ssc 0}$	&0 	&0 	&0	&0 	&  \multicolumn{2}{c||}{ $[K_i,K'_{i}]=-K''_{\ssc 0}$ for E}\\
$[J_{{\ssc 0}i},K'_{\ssc 0}]=$		&$-K'_{i}$	&0	&0	&$-K'_{i}$ 	&$-K'_{i}$ 	&0 	&0 	& \multicolumn{2}{c||}{ $[K_i,-K'_{\ssc 0}]=K''_i$ for C, D}\\
$[K'_{i},H']=$		&$P^\prime_{i}$	&$P^\prime_{i}$	&$P^\prime_{i}$	&$P^\prime_{i}$	&0 	&$P^\prime_{i}$ 	&0 	& \multicolumn{2}{c||}{$[K''_{i},H']=P_i$  for B, D}\\
& & \multicolumn{6}{c||}{ \big($H''=\frac{1}{c}H'$ for A, C, E, F \big)}  & \multicolumn{2}{c||}{$[K'_{i},H'']=P_i$  for E, F}\\

$[K'_{\ssc 0},H']=$		&$P^\prime_{\ssc 0}$	&0	&0	&$P^\prime_{\ssc 0}$	&0 	&0 	&0 	& \multicolumn{2}{c||}{$[K'_{\ssc 0}, H']=H$ for D}\\
\hline\hline
\end{tabular}
\end{center}
\vspace*{.2in}
\hrule
\end{table}

The important set of extra generators all the new symmetries have, apart from
the $G(0,3)$ subalgebra, is the $K'_i$ or $K''_i$. The latter transform as
components of a vector on the three dimensional space on which we have the
$SO(0,3)$ rotations, similar to the $K_i$ and $P_i$ vectors.
They commute with the $P_i$ translations in all cases, and commute also with
the $K_i$ velocity boosts in all except case E. For case A there is no further
extra nonzero commutator. So apart from the $[K_i,H]=P_i$ required in $G(0,3)$,
it has a structure similar to the static symmetry $S(0,3)$. Hence we named it
$G_S(0,3)$. This simple structure may have little to offer in a physics picture.
We can see that the $K''_i$ for cases B and D maintain the full algebraic structure
of a kind certaun of boosts on the 3D space with $H^\prime$ as the corresponding
`Hamiltonian', {\it i.e.} we have $[K''_{i},H']=P_i$. Case F is similar, with the
`Hamiltonian' being $H''$ and it has $K'_i$ instead of $K''_i$. Actually,
case F yields an algebra isomorphic to that of case B. Case D differs from
the latter in having two more nonzero commutators : $[K_i,-K'_{\ssc 0}]=K''_i$
and $[K'_{\ssc 0}, H']=H$. The former shows a relation between $K_i$ and $K''_i$
similar to that between $K_i$ and $P_i$ with the role of $-K'_{\ssc 0}$ to be
matched to that of $H$. So the  $K''_i$ are translations relative to the boosts
$K_i$ and boosts relative to the translations $P_i$. The structure is quite
complicated, somewhat like having a type of double layer  Galilean
structure among $K_i$, $K''_i$, and $P_i$. We denote the case B (and hence
also F) algebra by $G_B(0,3)$ for having an extra set of boosts, and that for
case D as $G_D(0,3)$ for the sort of double Galilean structure. $G_D(0,3)$ is
the one with the most complicated structure among all cases. However, it happens to
be the only one obtainable in a $Z_2^{\otimes \ssc N}$-graded contraction framework
\cite{ck} which includes all Cayley-Klein symmetries \cite{ckm}. From that
mathematical point of view, it looks like a natural candidate in the sequence :
\[
SO(2,4) \;\; \rightarrow  \;\; ISO(1,4)  \;\; \rightarrow
 \;\; G(1,3)  \;\; \rightarrow \;\;  G_D(0,3) \;.
\]
More details on that aspect we leave to the appendix.

For the remaining two cases, the $K'_{i}$ or $K''_i$ vector does not behave like
a set of boosts in terms of its relationship to any translation vector like $P_i$.
Case C has  $[K_i,-K'_{\ssc 0}]=K''_i$ as the  only nonzero commutation relation
beyond $J_{ij}$ and the three vector sets. So it has the $K''_i$ vector behaving
like the translations $P_i$, with the role of $-K'_{\ssc 0}$ being matched to that
of $H$ hence like another `Hamiltonian'. The algebra has an extra set of translations,
hence it is denoted by $G_T(0,3)$. Finally, we look at case E. The special commutation
relation $[K_i,K'_{i}]=-K''_{\ssc 0}$ may remind one of the nonzero commutator
between the two vector sets of generators as in the $C(m,n)$ case (see table~\ref{table1}).
Recall that $K'_{i}$ in this case behave like  boosts with respect
to $P_i$, in contrast to its behaving like translations in $C(0,3)$ with respect
to $K_i$.  We denote it by $G_C(0,3)$. In fact, we expect the symmetry to be
accessible via a contraction from $C(1,3)$.

The above description of contractions from $G(1,3)$ to the various three
dimensional relativity symmetries can obviously be generalized to all the
cases of any $G(m,n)$ giving the five new symmetries as $m+n-1$ dimensional
relativity symmetries with a $G(m-1,n)$ and $G(m,n-1)$ subgroup.

In the above, though we have been using the terms translations and boosts,
they refer only to the algebraic structure --- commutation relations with
other generators. The only geometric picture we stick to is that of
$SO(m,n)$ as rotations. In the next section, we look into some of the exact
physics pictures.

\section{Realization on the geometric arena}
As said, we think about a relativity symmetry as one that is the symmetry of
a classical geometric arena similar to but possibly beyond space(time), or
reference frame transformations thereof. Here we discuss the plausible
geometric picture of some of the new relativity symmetries introduced above.
We are interested on the physics side of the contraction pictures motivated by
Refs.\cite{023,030}.
The $SO(2,4)$ Quantum Relativity is formulated as a rotational type isometry
on a classical six-geometry with two non-spacetime coordinates ($u$-coordinates)
\cite{030}, giving a sort of description of a four dimensional noncommutative
spacetime. The relevant part of the six-geometry is only a five dimensional
hypersurface satisfying a constraint -- a `space-like AdS$_5$'. Some
description of the geometric picture has been given without any explicit
dynamical notion in the paper.

For the $ISO(1,4)$, named Snyder Relativity in Ref.\cite{036}, a quite
standard geometric picture of rotations and translations on a (classical)
five-geometry with the fifth coordinate being a $u$-coordinate has been adopted.
It is a five dimensional (geometric) space of Minkowski type with  the coordinate
$x^{\ssc 4}$ supplemented to the familiar Minkowski space $M^4$ of Einstein
spacetime, mathematically it is the natural coset space
\[
M^5 = ISO(1,4)/SO(1,4) \;.
\]
This fifth coordinate in the physics picture is to be written as
$x^{\ssc 4}=\bar{\kappa}\sigma$ with $\bar{\kappa}$
being essentially the parameter to be used in a further
contraction from $ISO(1,4)$, as given in Eqn.(\ref{kap}). The $\bar{\kappa}$
parameter is an  imposed invariant momentum, in the spirit of Snyder\cite{S}
essentially adopted from Ref.\cite{dsr} (see Refs.\cite{023,030}).
It assumes the physical dimension of momentum.
A further contraction gives the $G(1,3)$ Poincar\'e-Snyder Relativity
\cite{036,037}, which contains the familiar Poincar\'e symmetry of $ISO(1,3)$
as a preserved subgroup. The process splits out $x^{\ssc 4}$ from the
remaining four spacetime coordinates, casting $\sigma$ as an external
(absolute) parameter, similar mathematically to the Galilean time.
It has, however, the physical dimension of $\frac{time}{mass}$,
The original $SO(1,4)$ subgroup is contracted to another $ISO(1,3)$ subgroup
of $G(1,3)$, which contains an extra set of boost generators ---
boosts in relation to the Poincar\'e translations and a new `Hamiltonian'
generator ($K^\prime_\mu$, $P^\prime_\mu$, and $P^\prime_{\ssc 4}$ of
table 1, respectively). The set of four generators transform as
a vector in relation to the $3+1$ dimensional rotation $SO(1,3) \subset G(1,3)$,
and the transformations they generate, are called momentum boosts
\footnote{The momentum boosts are $\sigma$-dependent translations on $M^4$,
introduced in Ref.\cite{023} going from Poincar\'e symmetry towards the
construction of the Quantum Relativity.
}.
The contraction reduces
the coset space $M^5= ISO(1,4)/SO(1,4)$ to
\[
M^4\times I\!\!R = G(1,3)/ISO(1,3) \;,
\]
with the $ISO(1,3)$ factored out being the one with the momentum boosts.
The $M^4$ in the new coset space can be identified as the usual Minkowski
spacetime while $I\!\!R$ denotes a line of $\sigma$ values. The
above is just the analog of the spacetime picture in the contraction of Einstein
Relativity of $ISO(1,3)$ to the Galilean Relativity of $G(0,3)$.
Explicitly, we have the $G(1,3)$ realization as given by
\beqa
&& x'^\mu=  {\Lambda^{\mu}}_{\!\nu} x^\nu + p^\mu \sigma + A^\mu \;,
\nonumber \\
&& \sigma' = \sigma + S \;,
\label{g13act}
\eeqa
where ${\Lambda^{\mu}}_{\!\nu}$ is the Lorentz transformation and the
$+p^\mu \sigma$ part the momentum boosts with the Poincar\'e-Snyder momentum
given by $p^\mu= dx^\mu/d\sigma$.
The picture of the $G(1,3)$ realization can be taken to formulate a canonical
realization \cite{PP} as in Hamiltonian mechanics \cite{037} or a projective
representation \cite{gq} as in quantum mechanics \cite{036}, which are the
standard particle dynamic formulations for the corresponding Galilean case.
The success of such analyzes suggests the validity of the picture.
The phase space Hamiltonian mechanics picture has $\sigma$ as a formal
evolution parameter, generated by the $\sigma$-Hamiltonian $H^\prime$. It
is easy to see the single particle phase space is just the eight dimensional
coset space
\[
G(1,3)/[SO(1,3) \times  T_{\ssc H'}]\;,
\]
where $T_{\ssc H'}$ is the subgroup of $\sigma$-translation generated by
$H^\prime$. The natural canonical pair of phase space coordinates are
$(x^\mu, p_\mu)$.
\footnote{
The coadjoint orbits of Lie groups are natural candidates for symplectic manifolds
(see for example Ref.\cite{gq}), which provide the best framework for the
description of this kind of phase space.
}

Next, we want to follow the picture and see what happens with further
contractions of $G(1,3)$ to the new three dimensional relativity symmetries
we described above, or as listed in Table~\ref{table2}. In all these symmetries
as contractions from $G(1,3)$ we keep the Poincar\'e $ISO(1,3)$ subgroup
of the spacetime part contracted to $G(0,3)$ as the usual Galilean symmetry.
The contraction has a structure similar to that of Eqn.(\ref{kap}), with
the index 0 instead of 4 being singled-out from $J_{\mu\nu}$ and $P_\mu$
giving the rotations $J_{ij}$ and translations $P^\prime_i$. The
contraction transformation involves the invariant speed $c$ and gives
$K_i=\frac{1}{c} J_{{\ssc 0}i}$ and $P_i=\frac{1}{c} P^\prime_i$ [$P_i$ here is
not to be identified exactly as parts of the $P_\mu$ of Eqn.(\ref{i14})].
The generator $P^\prime_{\ssc 0}$ in the preserved subalgebra for all
cases of Table~(\ref{table2}) becomes the usual Hamiltonian that describes
time evolution, as the time $t$ from $x^{\ssc 0}= c\,t$ splits off as another
external parameter to the remaining three dimensional space. Naively, we
have an arena as $I\!\!R^3 \times I\!\!R \times I\!\!R$ described by
$(x^i, t, \sigma)$.

To start on a firm footing, we go back to the Newtonian space-time coset
representation of $G(0,3)$ as given by
\beqa
&& x^{\prime i}=  {R^{i}}_{j} x^j + V^i t + A^i \;,
\nonumber \\
&& t' = t + B \;.
\eeqa
For a generic element of the corresponding algebra, an infinitesimal
transformation, we have
\beq \label{g3}
\left(\begin{array}{c}
dt \\ dx^{i} \\    0
\end{array}\right)
=\left(\begin{array}{ccc}
{0} & {0} & b\\
v^i & \omega^i_j & a^i \\
 {0}   & {0} &  {0}
\end{array}\right)
\left(\begin{array}{c}
t \\ x^j \\    1
\end{array}\right)
=\left(\begin{array}{c}
 b\\ \omega^i_j  x^j +v^i t +a^i\\    0
\end{array}\right) \;,
\eeq
where $\omega^i_j, v^i, a^i, b$ denotes the set of (infinitesimal) parameters
to be exponentiated to describe the group manifold. The matching matrix
representation of an infinitesimal transformation the Poincar\'e $ISO(1,3)$
symmetry is given by
\beq \label{i13c}
\left(\begin{array}{c}
cdt \\ dx^{i} \\    0
\end{array}\right) =
\left(\begin{array}{ccc}
0 & \beta_j &  b \\
\beta^i & \omega^i_j  &  a^i/c \\
 {0}   & {0} &  0
\end{array}\right)
\left(\begin{array}{c}
ct \\x^j \\    c
\end{array}\right)
=\left(\begin{array}{c}
\beta_j  x^j + c b \\ \omega^i_j  x^j +\beta^i ct  +a^i\\   0
\end{array}\right) \;,
\eeq
where $\beta^i=v^i/c$. The expression of the above equation maybe in an
unfamiliar form. The notion involved is somewhat tricky, hence we will
walk the readers through it. Firstly, note that applying
$x^{\ssc 0}=ct$, $\omega^i_{\ssc 0}=\omega^{\ssc 0}_i=\beta_i$, and
$a^{\ssc 0}= c b$ give the equation in the form
\beq\label{i4c}
\left(\begin{array}{c}
dx^\mu \\ 0
\end{array}\right)
=\left(\begin{array}{cc}
\omega^\mu_\nu & a^\mu/c \\
0 & 0
\end{array}\right)
\left(\begin{array}{c}
x^\nu \\ c
\end{array}\right)
= \left(\begin{array}{c}
\omega^\mu_\nu x^\nu +  a^\mu \\ 0
\end{array}\right) \;.
\eeq
That is a version of the $ISO(1,3)/SO(1,3)$ coset space representation for the
Poincar\'e symmetry.
One can also check that Eqn.(\ref{i13c}) goes with the
$ISO(1,3)\rightarrow G(0,3)$ contraction to exactly Eqn.(\ref{g3}). Recall
from the discussion in the previous section that in the contraction, one
first rewrites the generators $J_{{\ssc 0}i}$ and $P^\prime_i$ of $ISO(1,3)$
as $K_i=\frac{1}{c} J_{{\ssc 0}i}$ and $P_i=\frac{1}{c} P^\prime_i$. Note
that in all the matrix representations of the transformations used here,
we have matrix elements of different physical dimensions. The elements as
parameters in the infinitesimal transformations, in particular, have to be
matched to their corresponding generators to give the same consistent physical
quantities to be summed  (with a quantity of the right dimension) as the
exponent for a group element. For example, we check that for the matrix
elements in Eqn.(\ref{g3}), we have quantities $J_{ij} \omega^i_j$, $K_i v^i$,
$P_i a^i$, and $H b$ all having the same dimension.
Working on inverting the contraction from Eqn.(\ref{i13c}) to Eqn.(\ref{g3}),
we focus first on the part other than the translations. As generators 
for $ISO(1,3)$, the $J_{{\ssc 0}i}$ need to have the same dimension
as $J_{ij}$, which is satisfied by $J_{{\ssc 0}i}= c K_i$. The parameter to
match $J_{{\ssc 0}i}$, $\omega^i_{\ssc 0}=\beta^i$, is naturally given by
$\frac{1}{c} v^i$. Consistency in the representation form is restored by
shifting the $c$ factor in $v^i =c \beta^i$ from the matrix to the vector
it acts on to form $ct$, as in Eqn.(\ref{i13c}). The contraction taking
Eqn.(\ref{i13c}) to Eqn.(\ref{g3}) has the limit of $c \to \infty$
taken with $\beta^i c$ kept constant, hence only the zero $\beta_j x^j$
drops out (to be exact, $\beta_j x^j/c$ drops out from $dt$). Similar
reasoning applies to the translational part, since the
spacial translation generators $P^\prime_i$ of $ISO(1,3)$ are $c$ times
the translation generators $P_i$ of $G(0,3)$. The corresponding parameters
should be $\frac{1}{c} a^i$ and a factor $c$ has to be multiplied to the
$1$ in the vector. The resulting $c$ actually gives the right quantity
for $cdt$, as it is multiplied by $b$. The parameters $a^i$ are kept constant
when taking the $c \to \infty$ contraction limit of Eqn.(\ref{i13c}). When
the $c$ factor in $cdt$ and $ct$ is taken out, the last element of the
vector is restored to $1$, as in Eqn.(\ref{g3}). So, to
rewrite Eqn.(\ref{i4c}) in the familiar form, the $P^\prime_\mu$
generators of $ISO(1,3)$ are actually replaced by $P_\mu = c P^\prime_\mu$.
The latter have to be handled with care when the whole $G(1,3)$ group
is considered.

Now, we can extend the above to look into the contraction of $G(1,3)$ in
the $G(1,3)/ISO(1,3)$ coset representation. We write the matrix form of the
infinitesimal transformation as
\beq \label{g13}
\left(\begin{array}{c}
cdt \\ dx^{i}  \\ cd\sigma \\     0
\end{array}\right)
=\left(\begin{array}{cccc}
 0 & v_j/c & p^{\ssc 0}/c & c b \\
v^i/c & \omega^i_j & p^i/c  & a^i\\
{0} & {0} & {0} & cs\\
 {0} & {0}   & {0} &  {0}
\end{array}\right)
\left(\begin{array}{c}
ct \\ x^j \\ c\sigma \\     1
\end{array}\right)
=\left(\begin{array}{c}
v_j  x^j/c + p^{\ssc 0} \sigma + c b \\
\omega^i_j  x^j +v^i t + p^i \sigma +a^i\\
 cs\\     0
\end{array}\right) \;,
\eeq
in which $P_\mu = \frac{1}{c} P^\prime_\mu$ are directly represented with
parameters $a^\mu$, as discussed above. Besides those in the Poincar\'e
subalgebra, we have the extra generators $K^\prime_\mu$ and $H^\prime$
with parameters here taken as $p^\mu/c$, and $s$, respectively. However,
$\frac{1}{c} H^\prime$ instead of $H^\prime$ is directly represented, hence
the parameter $cs$. Note that the product $H^\prime K^\prime_\mu$ should
have the same dimension as that of $P^\prime_\mu$, hence $sp^\mu/c$ that
of $a^\mu/c$. The special choices are what is needed to go from the more
natural form as an extension of Eqn.(\ref{i13c}) or Eqn.(\ref{i4c}) to one in
the form Eqn.(\ref{g3}) instead. The contracted symmetries $G_S(0,3)$,
$G_B(0,3)$, $G_T(0,3)$, and $G_D(0,3)$, all have $K^"_i=\frac{1}{c}K^\prime_i$,
which have $p^i$ as parameters. In the case of $G_S(0,3)$ and $G_B(0,3)$,
we have also $K^"_{\ssc 0}=\frac{1}{c}K^\prime_{\ssc 0}$ with parameter
$p^{\ssc 0}$. Furthermore, we have to represent the preserved $H^\prime$.
We can rewrite Eqn.(\ref{g13}) for this case as
\beq \label{g13b}
\left(\begin{array}{c}
cdt \\ dx^{i}  \\ d\sigma \\     0
\end{array}\right)
=\left(\begin{array}{cccc}
 0 & v_j/c & p^{\ssc 0} & c b \\
v^i/c & \omega^i_j & p^i  & a^i\\
{0} & {0} & {0} & s\\
 {0} & {0}   & {0} &  {0}
\end{array}\right)
\left(\begin{array}{c}
ct \\ x^j \\ \sigma \\     1
\end{array}\right)
=\left(\begin{array}{c}
v_j  x^j/c + p^{\ssc 0} \sigma + c b \\
\omega^i_j  x^j +v^i t + p^i \sigma +a^i\\
 s\\     0
\end{array}\right) \;.
\eeq
Taking $c \to \infty$ then leaves \\
-- $G_B(0,3)$  :
\beq \label{gb}
\left(\begin{array}{c}
dt \\ dx^{i}  \\ d\sigma \\     0
\end{array}\right)
=\left(\begin{array}{cccc}
 0 & 0 & 0 &  b \\
v^i & \omega^i_j & p^i  & a^i\\
{0} & {0} & {0} & s\\
 {0} & {0}   & {0} &  {0}
\end{array}\right)
\left(\begin{array}{c}
t \\ x^j \\ \sigma \\     1
\end{array}\right)
=\left(\begin{array}{c}
 b \\
\omega^i_j  x^j +v^i t + p^i \sigma +a^i\\
 s\\     0
\end{array}\right) \;.
\eeq
For the case of $G_S(0,3)$, we have to further replace $H^\prime$ by
$H^"=\frac{1}{c} H^\prime$ with parameter $\bar{s}=cs$ characterizing
translations in $\bar{\sigma}=c\sigma$. At the $c \to \infty$ limit,
we have\\
-- $G_S(0,3)$  :
\beq \label{gs}
\left(\begin{array}{c}
dt \\ dx^{i}  \\ d\bar{\sigma} \\     0
\end{array}\right)
=\left(\begin{array}{cccc}
 0 & 0 & 0 &  b \\
v^i & \omega^i_j & 0  & a^i\\
{0} & {0} & {0} &  \bar{s}\\
 {0} & {0}   & {0} &  {0}
\end{array}\right)
\left(\begin{array}{c}
t \\ x^j \\ \bar{\sigma} \\     1
\end{array}\right)
=\left(\begin{array}{c}
 b \\
\omega^i_j  x^j +v^i t  +a^i\\
 \bar{s}\\     0
\end{array}\right) \;.
\eeq
We have illustrated how to obtain the results for the cases of $G_B(0,3)$ and
$G_S(0,3)$. We skip further details and list the results for the rest of the
cases. \\
-- $G_D(0,3)$  ($\rho^{\ssc 0}=p^{\ssc 0}/c$) :
\beq \label{gd}
\left(\begin{array}{c}
dt \\ dx^{i}  \\ d\sigma \\     0
\end{array}\right)
=\left(\begin{array}{cccc}
 0 & 0 & \rho^{\ssc 0} &  b \\
v^i & \omega^i_j & p^i  & a^i\\
{0} & {0} & {0} & s\\
 {0} & {0}   & {0} &  {0}
\end{array}\right)
\left(\begin{array}{c}
t \\ x^j \\ \sigma \\     1
\end{array}\right)
=\left(\begin{array}{c}
 \rho^{\ssc 0} \sigma + b \\
\omega^i_j  x^j +v^i t + p^i \sigma +a^i\\
 s\\     0
\end{array}\right) \;;
\eeq
-- $G_C(0,3)$ ($\rho^i=p^i/c$) :
\beq \label{gc}
\left(\begin{array}{c}
dt \\ dx^{i}  \\ d\bar{\sigma} \\     0
\end{array}\right)
=\left(\begin{array}{cccc}
 0 & 0 & 0 &  b \\
v^i & \omega^i_j & \rho^i  & a^i\\
{0} & {0} & {0} & \bar{s}\\
 {0} & {0}   & {0} &  {0}
\end{array}\right)
\left(\begin{array}{c}
t \\ x^j \\ \bar{\sigma} \\     1
\end{array}\right)
=\left(\begin{array}{c}
   b \\
\omega^i_j  x^j +v^i t + \rho^i \bar{\sigma} +a^i\\
 \bar{s}\\     0
\end{array}\right) \;;
\eeq
-- $G_T(0,3)$ the same as $G_S(0,3)$ ;\\
-- $G_B(0,3)$ for case F the same as  $G_C(0,3)$ .\\
We include the last case for $G_B(0,3)$ as given under the column of
case F in Table~\ref{table2} for completeness. It is of course equivalent
to that of case B given above, only given explicitly in $\rho^i$ and
$\bar{\sigma}$ instead of of $p^i$ and $\sigma$. That explains the apparent
different contractions leading to the two $G_B(0,3)$ results.

We have analyzed above the contractions of the $G(1,3)$ representation
given by Eqn.(\ref{g13act}) explicitly. The five different contracted
symmetries of Table~\ref{table2} give only three inequivalent results.
The representation arena obviously has the geometry
\[
I\!\!R^3 \times I\!\!R \times I\!\!R_* \;,
\]
where $I\!\!R$ denotes a line of $t$ values and $I\!\!R_*$ one of $\sigma$
or $\bar{\sigma}$ values. The geometry can be considered as a coset space,
which is for each of the cases explicitly given as :
\beqa
G_S(0,3) &---& G_S(0,3)/S(0,3) \;,
\nonumber \\
G_B(0,3) &---& G_B(0,3)/S(0,3) \;,
\nonumber \\
G_D(0,3) &---& G_D(0,3)/G(0,3) \;,
\nonumber \\
G_T(0,3) &---& G_T(0,3)/G(0,3) \;,
\nonumber \\
G_C(0,3) &---& G_C(0,3)/C(0,3) \;.
\nonumber \eeqa
The results can also be understood from the algebraic structure discussed
in the previous section. First note that the coset representations of this
type singles out at a vector set of generators of the translation kind to
describe the basic part of the physics arena, here the isotropic Newtonian
space $I\!\!R^3$. The translations commute with all others generators
except the rotations. The Galilean boost vector and the Hamiltonian, with
the characterizing commutators $[K_i, H]=P_i$ asks for the time dimension.
In the simplest case of the $G_S(0,3)$ and $G_T(0,3)$ groups, no commutator
between each of the extra five generators beyond those of the $G(0,3)$
subgroup and any other generator (shown in the lower part of Table~\ref{table2})
involves the translations $P_i$ or $H$. Hence, the extra symmetry transformations
are trivialized on the corresponding cosets, which are equivalent. The extra
coordinate $\bar{\sigma}$ as a remnant from the $G(1,3)/ISO(1,3)$ coset is
irrelevant as it has no role to play in any particle dynamics on the
$I\!\!R^3$ [{\it cf.} Eqn.(\ref{gs})]. In the cases of $G_B(0,3)$ and
$G_C(0,3)$, $P_i$ shows up in one equivalent set of commutators defining
an extra boost vector relative to the $P_i$. Again, the difference in the
commutators not involving the $P_i$ at all has no implication on the coset
representations. Although $G_C(0,3)$ is a more complicated group, the
corresponding coset representation is equivalent to that of $G_B(0,3)$,
showing a second boost set with another coordinate similar to time
[{\it cf.} Eqn.(\ref{gb})]. Finally, the structurally most complicated
$G_D(0,3)$ seen as action on the $P_i$ has also simply extra boosts,
with the only extra structure nontrivially realized being the commutator
$[K^\prime_{\ssc 0}, H^\prime]=H$. One can interpret the latter as
$K^\prime_{\ssc 0}$ being a one dimensional boost with respective to
the Hamiltonian $H^\prime$ on the space of the translation $H$, which
shows in the first row of the matrix Eqn.(\ref{gd}).

\section{Conclusions}
We have discussed in this paper the perspective of an $N$ dimensional
relativity symmetry having at least an $SO(m,n)$ subgroup with $N=m+n$
as rotations of an $N$ dimensional arena of classical geometry, obtainable
from contractions of rotational symmetry on a higher dimensional geometry.
We focused mostly on simple In\"on\"u-Wigner contractions that reduces the
dimension of the relativity by one at a time. We have summarized the generic
relativity symmetries at $N-2$ dimensions as defined obtainable from the
contractions. As the most interesting part of the results, we have presented
the five relativity symmetries at $N-3$ dimensions obtainable from one
further contraction of Galilean type symmetries at the $N-2$ level
keeping the $N-3$ Galilean subgroup, using the example of
$G(0,3) \subset G(1,3)$. For the latter case, we have presented the corresponding 
coset space representation along the lines of the usual Newtonian space-time
or Einstein/Minkowski spacetime picture. The latter is essentially
singled out by our basic perspective of a relativity symmetry as the
basic arena for the physics of the relativity. The above gives concrete
settings for the description of dynamics.

The first thing readers will realize is the very rich set of options
in contrast to the simple results at the $N-1$ and $N-2$ levels. However,
the sequence exemplified by
\[
SO(2,4) \;\; \rightarrow  \;\; ISO(1,4)  \;\; \rightarrow
 \;\; G(1,3)  \;\; \rightarrow \;\;  G_D(0,3) \;
\]
seems to single itself out from various point of view. This particular
example, maybe with some other groups containing $G(0,3)$ replacing
$G_D(0,3)$, is mostly what we have been motivated to study from our
project on Quantum Relativity. That fits in well with our
analysis of $G(1,3)$ Poincar\'e-Snyder mechanics.
$G_D(0,3)$ is quite complicated,
but the corresponding coset representation picture is very manageable.
We would like to follow up on the physics of the dynamics.
We believe the analysis in the paper also has some interesting
mathematical results.

\appendix{ }
\section{ }
On the mathematical side 1991 saw the introduction of the generalized
(In\"on\"u-Wigner) contraction by Weimar-Woods \cite{glc} and the
graded contractions from the purely algebraic perspective \cite{gdc}.
Studies of the topics are still active, with also extensions beyond the Lie
algebra setting \cite{BHRS}. The generalized contraction allows the generators to
be scaled with different powers of the contraction parameter. It can describe
the combined effect of repeated applications of simple contractions of
the In\"on\"u-Wigner type. An example is given by $SO(1,4)\rightarrow G(0,3)$
through the $\epsilon \to 0$ limit of $K_i= \epsilon J_{{\ssc 0}i}$,
$H= \epsilon J_{{\ssc 04}}$, and $P_i= \epsilon^2 J_{i{\ssc 4}}$. The
$\epsilon$ parameter serves here both as the $1/\bar{\kappa}$ and
$1/c$ in the explicit example of the
$ISO(1,4) \rightarrow  G(1,3) \rightarrow G_D(0,3)$ contractions
discussed in the paper looking only at the subgroup part of
$SO(1,4) \rightarrow ISO(1,3) \rightarrow G(0,3)$. So long as the
mathematical structure is concerned, the contraction parameter is only
a tool and only the singular limit is important. It is easy to see,
however, that in a physical application, the meaning of the contraction
parameter is of paramount importance. Hence, combining repeated applications
of In\"on\"u-Wigner contractions may not be an advantage. Similarly, while
the formulation of graded contractions is of great value in mathematics,
its direct application in physics is likely to be very limited. In our
opinion, it will likely serve as a tool to identify interesting
contracted symmetries to be studied as the end product of a sequence of
In\"on\"u-Wigner contractions.

\begin{table}[t]
\caption{Some Cayley-Klein algebras from  $Z_2^{\otimes \ssc N}$-graded
contractions of $SO(N+1)$. Note that in all cases, $m+n=[N+1 -(\mbox{\# of zeros})]$
gives the relativity symmetry dimension.}
\label{CK}
\begin{center}
\begin{tabular}{|c|c|c|c|}\hline\hline
$(\kappa_{1},\kappa_{2},\cdots,\kappa_{N})$  &\# of zeros	&\# of nonzero commutators	& symmetry\\ \hline\hline
$(\pm 1,\cdots,\pm 1)$		&0	&$\frac{1}{2}(N+1)N(N-1)$	&$SO(m,n)$\\\hline
$(0,\pm 1,\cdots,\pm 1)$,$(\pm 1,\cdots,\pm 1,0)$		&1	&$\frac{1}{2}N^2(N-1)$	&$ISO(m,n)$\\\hline
$(\pm 1,0,\pm 1,\cdots,\pm 1)$,$(\pm 1,\cdots,\pm 1,0,\pm 1)$		&1	&$\frac{1}{2}(N-1)(N^2-N+2)$	& $NH^\pm (m,n)$\\\hline
$(0,0,\pm 1,\cdots,\pm 1)$,$(\pm 1,\cdots,\pm 1,0,0)$		&2	&$\frac{1}{2}N(N-1)^2$	&$G(m,n)$\\\hline
$(0,\pm 1,\cdots,\pm 1,0)$		
&2	&$\frac{1}{2}N(N-1)^2$	&$C(m,n)$\\\hline
$(0,0,0,\pm 1,\cdots,\pm 1)$,$(\pm 1,\cdots,\pm 1,0,0,0)$		&3	&$\frac{1}{2}(N^3-3N^2+2N+2)$	&$G_{D}(m,n)$\\ \hline\hline
\end{tabular}
\end{center}
\end{table}

Graded contraction, especially in the case of the
$Z_2^{\otimes \ssc N}$-graded contraction, is different enough
in formulation and interesting enough in some of the patterns shown that
we want to give the description of the relativity symmetries we
discussed in that picture. For a particularly interesting set of the
$Z_2^{\otimes \ssc N}$-graded contraction on an $SO(N+1)$ algebra
\cite{ck}, the nonzero commutators are written as
\beq
[J_{ab},J_{ac}]=\kappa_{ab} J_{bc} \;, \quad
[J_{ab},J_{bc}]=- J_{ac}\;, \quad
[J_{ac},J_{bc}]=\kappa_{bc} J_{ab}\;,  \qquad
(a<b<c),
\eeq
where
\beq
\kappa_{ab}:=\kappa_{a+1}\kappa_{a+2}\cdots\kappa_{b}  \qquad
(a,b=0,1,2,\cdots,N,a<b),
\eeq
and the set of $\kappa_{a}$ values fixes a particular contraction. The set are
all orthogonal Cayley-Klein algebras are the motion algebras of the geometries of a 
real space with a projective metric \cite{ckm}. Of course indexing of the generators
of the $SO(N+1)$ can be shuffled before fitting into the scheme. Without loss
of generality, one can consider only the values of $\pm 1$ and 0 for each
$\kappa_{a}$. All $\kappa_{a}$ being $+1$ gives the original $SO(N+1)$. Having
some $\kappa_{a}$ being $-1$ corresponds to a different real form of the same
complex algebra as the one without negative $\kappa_{a}$ values. Having some
$\kappa_{a}$ being zero corresponds to a real generalized contraction in the
In\"on\"u-Wigner sense of the singular limit. We give the cases of our interest
in Table~\ref{CK}. One can see some interesting patterns, mostly obvious from
the table. Firstly, a sub-sequence $\pm 1$ indicates an $SO(m,n)$ subalgebra.
In fact, a particular sub-sequence also indicates the corresponding subalgebra.
A single In\"on\"u-Wigner reducing the relativity symmetry dimension by one
imposes a zero from one end of the core sub-sequence for the $SO(m,n)$
subalgebra, one reducing the dimension by $n$ directly would impose a zero
at the $n$-th position from one end of the core sub-sequence. A zero adjacent
to an $SO(m,n)$ sequence indicates a vector set of symmetries like
translations. A further adjacent zero indicates a vector set as boosts
relative to the first set as translations. For three adjacent zeros like
the case of $G_{D}(m,n)$, we have three vectors, the first is translations
relative to the second as boosts while the second is translations
relative to the third as boosts. Hence, $C(m,n)$ has two set of
vectors without any translation-boost pair. Similarly, the Newton-Hooke
symmetries $NH^+(m,n)$ or $NH^-(m,n)$ has respectively an $SO(2)$ and
an $SO(1,1)$ subalgebra, and indicators for the $+1$ and $-1$ of the isolated
nonzero $\kappa_{a}$, in addition to the $SO(m,n)$. The other generators
form a $2\times (m+n)$ set as two sets of $m+n$ vectors and $m+n$ sets of
two-vectors. The scheme does not include all algebras accessible through
contractions of an $SO(m,n)$. A clear example is given by the $S(m,n)$.

Let us check a bit the contractions from $G(m,n)$ as discussed in the main
text. In the example of contractions of $G(1,3)$ preserving a $G(0,3)$, one
should be looking into a five $\kappa_{a}$ sequence with a $(0,0,1,1)$
sub-sequence. The only inequivalent options are $(0,0,0,1,1)$, $(1,0,0,1,1)$,
and $(0,0,1,1,0)$. The first is the $G_D(0,3)$ shown, and the last has the
a $C(0,3)$ subalgebra, which does not fit the $G_C(0,3)$ either. Among the
two vector sets of generators in the $G(0,3)$ subalgebra, the one that joins
the extra vector set to the $C(0,3)$ subalgebra in $(0,0,1,1,0)$ is the
translations. For the $G_C(0,3)$, it is rather the boosts. For the algebra of
$(1,0,0,1,1)$, there is no three vector set of generators with respect to
the $SO(0,3)$, hence it does not fit any of the symmetries discussed in the
text which preserve the third vector set. In fact, the $(1,0,0,1,1)$
structure suggests that it cannot be obtained from contraction of the
$(0,0,1,1,1)$ structure of $G(1,3)$. It is also clear that the other
symmetries besides $G_D(0,3)$ given in Table~\ref{table2} are not included
in the scheme summarized here in the appendix. It is easy to see that the
conclusions also generalize to replacing the $(0,3)$ with any $(m,n)$.

\vspace*{.2in}
\noindent{\em Acknowledgements :-\ }
We thank  Hung-Yi Lee for discussions and J. Nester for
helping to improve the English presentation.
The work is partially supported by  the research grant
No. 99-2112-M-008-003-MY3 from the NSC of Taiwan.

\end{document}